\documentstyle[12pt]{article}

\newcommand{\beq}{\begin{equation}}
\newcommand{\eeq}{\end{equation}}
\newcommand{\ba}{\begin{array}}
\newcommand{\ea}{\end{array}}
\newcommand{\bea}{\begin{eqnarray}}
\newcommand{\eea}{\end{eqnarray}}
\newcommand{\bean}{\begin{eqnarray*}}
\newcommand{\eean}{\end{eqnarray*}}

\newcommand{\nin}{\noindent}
\newcommand{\ra}{\rightarrow}
\newcommand{\ee}{{\rm e}}
\newcommand{\eee}{{\bf e}}
\newcommand{\fff}{{\bf f}}

\makeatletter \ifcase\@ptsize \font\teneufm=eufm10
\font\seveneufm=eufm7 \font\fiveeufm=eufm5
\font\teneusm=eusm10 \font\seveneusm=eusm7
\font\fiveeusm=eusm5 \or \font\teneufm=eufm10 scaled
\magstephalf \font\seveneufm=eufm7 \font\fiveeufm=eufm5
\font\teneusm=eusm10 scaled \magstephalf
\font\seveneusm=eusm7 \font\fiveeusm=eusm5 \or
\font\teneufm=eufm10 scaled \magstep1 \font\seveneufm=eufm7
\font\fiveeufm=eufm5 \font\teneusm=eusm10 scaled \magstep1
\font\seveneusm=eusm7 \font\fiveeusm=eusm5 \fi

\newfam\eufmfam \newfam\eusmfam \textfont\eufmfam=\teneufm
\scriptfont\eufmfam=\seveneufm
\scriptscriptfont\eufmfam=\fiveeufm
\textfont\eusmfam=\teneusm \scriptfont\eusmfam=\seveneusm
\scriptscriptfont\eusmfam=\fiveeusm

\def\frak{\ifmmode\let\next\frak@\else
 \def\next{\errmessage{Use \string\frak\space only in math
 mode}}\fi\next} \def\frak@#1{{\frak@@{#1}}}
 \def\frak@@#1{\fam\eufmfam#1} 
 \def\sh{\ifmmode\let\next\sh@\else
 \def\next{\errmessage{Use \string\sh\space only in math
 mode}}\fi\next} \def\sh@#1{{\sh@@{#1}}}
 \def\sh@@#1{\fam\eusmfam#1}

\ifcase\@ptsize \font\tenmsa=msam10 \font\sevenmsa=msam7
 \font\fivemsa=msam5 \font\tenmsb=msbm10
 \font\sevenmsb=msbm7 \font\fivemsb=msbm5 \or
 \font\tenmsa=msam10 scaled \magstephalf
 \font\sevenmsa=msam7 \font\fivemsa=msam5
 \font\tenmsb=msbm10 scaled \magstephalf
 \font\sevenmsb=msbm7 \font\fivemsb=msbm5 \or
 \font\tenmsa=msam10 scaled \magstep1 \font\sevenmsa=msam7
 \font\fivemsa=msam5 \font\tenmsb=msbm10 scaled \magstep1
 \font\sevenmsb=msbm7 \font\fivemsb=msbm5 \fi

\newfam\msafam \newfam\msbfam \textfont\msafam=\tenmsa
\scriptfont\msafam=\sevenmsa
\scriptscriptfont\msafam=\fivemsa \textfont\msbfam=\tenmsb
\scriptfont\msbfam=\sevenmsb
\scriptscriptfont\msbfam=\fivemsb

\def\Bbb{\ifmmode\let\next\Bbb@\else
 \def\next{\errmessage{Use \string\Bbb\space only in math
 mode}}\fi\next} \def\Bbb@#1{{\Bbb@@{#1}}}
 \def\Bbb@@#1{\fam\msbfam#1} \def\hexnumber@#1{\ifnum#1<10
 \number#1\else \ifnum#1=10 A\else\ifnum#1=11
 B\else\ifnum#1=12 C\else \ifnum#1=13 D\else\ifnum#1=14
 E\else\ifnum#1=15 F\fi\fi\fi\fi\fi\fi\fi}
 \def\msa@{\hexnumber@\msafam} \def\msb@{\hexnumber@\msbfam}
 \mathchardef\square="0\msa@03

\makeatother
\newcommand{\HH}{{\Bbb H}} \newcommand{\RR}{{\Bbb R}}
\newcommand{\CC}{{\Bbb C}} \newcommand{\PP}{{\Bbb P}}
 \newcommand{\GG}{{\Bbb G}}
\newcommand{\SS}{{\Bbb S}} 
\newcommand{\EE}{{\Bbb E}}
\begin{document}

\title{Holomorphic curves and Toda systems\thanks{This
work was supported by KBN grant 2P03B18509}}
\author{Adam Doliwa \\
Institute of Theoretical Physics, Warsaw University \\
ul. Ho\.{z}a 69, 00-681 Warsaw, Poland \\
e-mail: doliwa@fuw.edu.pl }
\date{}
\maketitle

\begin{abstract}
\noindent Geometry of holomorphic curves from point of view of open Toda
systems is discussed. Parametrization of curves related this way to
non-exceptional simple Lie algebras is given. This gives rise to explicit
formulas for minimal surfaces in real, complex and quaternionic projective
spaces or complex quadrics. The paper
generalizes the well known connection between minimal surfaces in $\EE^{3}$,
their Weierstrass representation in terms of holomorphic functions and the
general solution to the Liouville equation.
\end{abstract}

\vskip4cm

\hbox to 5truein{ \hfil {\large {\bf Preprint IFT 7/95}}}

\vskip4cm

\pagebreak

\section{Introduction}

\label{sec:Introduction}

Let ${\frak g}$ be a simple complex Lie algebra, ${\frak h}$ its Cartan
subalgebra, $\Delta \subset {\frak h}^{*}$ the root system and
\beq {\frak g} = {\frak h} + \sum_{\alpha\in\Delta}{\frak g}_{\alpha} \eeq
the corresponding root space decomposition (see e.g. \cite{Hump}). We denote by
$\pi\subset\Delta$ a system of simple roots $\pi =\{ \alpha_{k}\}_{k=1}^{m}$
($m=\dim {\frak h}$) and by $\{ h_{k} \}_{k=1}^{m}$
the basis of ${\frak h}$ dual to $\{ \alpha_{k}\}_{k=1}^{m}$ with respect to
the Killing form of ${\frak g}$.

\bigskip

\nin The open Toda system related to ${\frak g}$ \cite{MOP} is the following
equation
\beq \label{eq:Toda}
2\theta_{,\xi\eta} = \sum_{k=1}^{m}h_{k}\ee^{-2\alpha_{k}(\theta)} \; \; ,
\eeq
where
\beq \label{def:theta}
\theta: \CC^{2}\ni(\xi,\eta) \mapsto \theta(\xi,\eta) \in {\frak h} \; \; \eeq
is unknown function with values in the Cartan subalgebra.

\bigskip

\nin In this paper we are interested in a particular reduction of the system
(\ref{eq:Toda})
\beq \xi=z \; \; \; , \; \; \; \eta =\bar{z} \; \; \; , \; \; \; \theta\in
\sum_{k=1}^{m}\RR h_{k} \; \;  \eeq
obtained from restriction of ${\frak g}$ to its compact real form.

\bigskip

\nin General solutions to the open Toda systems were found in \cite{LeSav}
using the theory of representation of Lie algebras. Recently appeared
papers \cite{GervMats,DolSy2,GervSav,SavRaz} where connection of the open Toda
systems to geometry of holomorphic curves was pointed out. This approach is
developed in the present Letter. The geometric nature of solutions
to the Toda systems, as coming from parametrization of the corresponding
holomorphic curves is shown.

\bigskip

\nin Out of such holomorphic curves one can construct minimal
surfaces (or more generally, harmonic maps) into
complex projective spaces, complex quadrics, Euclidean spheres and quaternionic
projective spaces (see \cite{Yang} and the references given there).
The method presented below can be considered as generalization (suggested in
\cite{DolSy1}) of the well
known connection between Weierstrass representation of minimal surfaces in
$\EE^{3}$ in terms of holomorphic
functions, and the simplest Toda system -- the Liouville equation.

\bigskip

\nin The next four Sections are devoted to presentation of the approach to
the Toda systems related to four classical sequences of simple Lie
algebras.

\section{Holomorphic curves and $A_{n}$-Toda systems}

 \label{sec:A}

Let $\phi: {\cal R} \ra \CC\PP^{n}$ be a nondegenerate  (i.e. not contained in
a proper projective subspace of $\CC\PP^{n}$) holomorphic curve
\cite{GrHa} represented locally by meromorphic vector-function
$f: {\cal R}\supset {\cal O} \ra \CC^{n+1}$. By $\phi_{k}: {\cal R} \ra
\GG(k,n+1)$ ($k=1,\ldots ,n$) we denote its associated curves ($\phi_{1}=\phi$)
represented locally by $k$-plane with matrix homogeneous coordinates
\beq
M_{k} = (f,f', \ldots ,f^{(k-1)} ) \in \CC^{n+1}_{k} \; \; . \eeq
Each of the curves $\phi_{k}$ induces on the Riemann surface ${\cal R}$ the
K\"{a}hler form $\omega_{k}$ (we make use the standard K\"{a}hler structure on
Grassman manifolds) locally expressed as
\beq
\omega_{k|{\cal O}}=
\frac{i}{2}\partial\bar{\partial} \log \det M_{k}^{+}M_{k} =
\frac{i}{2}\partial\bar{\partial} \log ||\Lambda_{k}||^{2} \; \; . \eeq
All the resulting Riemannian metrics on ${\cal R}$ are conformally equivalent,
moreover they are subjected to the so called Pl\"{u}cker formulas
\beq \label{eq:Plucker}
\frac{\partial^{2}}{\partial z \partial\bar{z}} \log ||\Lambda_{k}(z)||^{2}  =
\frac{||\Lambda_{k-1}(z)||^{2}
||\Lambda_{k+1}(z)||^{2}}{||\Lambda_{k}(z)||^{4}}
 \; \; , \; \; k=1, \ldots , n
\eeq
where $||\Lambda_{0}||^{2} = 1$ and $||\Lambda_{n+1}||^{2} =
|\det M_{n+1}|^{2}$.

\bigskip

\nin When the local holomorphic lift $f$ of the curve $\phi$ is subjected to
the
normalization condition
\beq \det M_{n+1} = \det (f,f', \ldots ,f^{(n)}) = 1 \eeq
then one can rewrite \cite{GervMats} the Pl\"{u}cker formulas
(\ref{eq:Plucker}) in one of the standard forms of the open Toda system
related to the ${\frak s}{\frak u}(n+1)$ Lie algebra (the compact real form of
$A_{n}\cong{\frak s}{\frak l}(n+1)$)
\beq \varphi^{k}_{,z\bar{z}} = \exp \left( -\sum_{l=1}^{n}C_{kl}\varphi^{l}
\right) \; \; , \; \; k=1,\ldots ,n \eeq
where
\beq \varphi_{k}=\log ||\Lambda_{k}||^{2} \eeq
and $C_{kl}$ is the Cartan matrix of the $A_{n}$ Lie algebra \cite{Hump}.

\medskip

\nin {\bf Theorem 1 \cite{GervMats}} The Pl\"{u}cker formulas for holomorphic
curves in $\CC\PP^{n}$ form the open Toda system
related to the compact real form of ${\frak s}{\frak l}(n+1)$ Lie algebra.

\bigskip

\nin For our purposes it is more convenient to rewrite the system
(\ref{eq:Plucker}) using functions
\beq \label{def:thetaA}
\theta^{k} = \log\frac{||\Lambda_{k}||}{||\Lambda_{k-1}||} \; \; \; , \;
\; k=1,\ldots ,n+1  \; \; \; , \; \; \sum_{k=1}^{n+1}\theta^{k} = 0 \; \; ,
\eeq
then
\bea \label{eq:TodaA}
2\theta^{1}_{,z\bar{z}} & = & \ee^{2(\theta^{2} - \theta^{1})} \nonumber \\
2\theta^{l}_{,z\bar{z}} & = & \ee^{2(\theta^{l+1} - \theta^{l})} -
 \ee^{2(\theta^{l} - \theta^{l-1})} \; \; , \; \; l=2, \ldots , n\nonumber \\
2\theta^{n+1}_{,z\bar{z}} & = & -\ee^{2(\theta^{n+1} - \theta^{n})} \; \; .
\eea
This matches with the standard representation of the simple roots and Cartan
subalgebra of $A_{n}$ (we use the scaled Killing form) in terms of the
orthonormal basis $\{\eee_{k}\}_{k=1}^{n+1}$ of of $\RR^{n+1}$
(see equation (\ref{eq:Toda}))
\beq
\pi = \{ \eee_{1}-\eee_{2},\eee_{2}-\eee_{3}, \ldots , \eee_{n}-\eee_{n+1} \}
\cong \{ h_{1}, h_{2}, \ldots ,h_{n} \} \; \; .
\eeq
Using Gram-Schmidt orthonormalization procedure one can obtain from the
natural basis $f,f',\ldots ,f^{(n)}$ along the nondegenerate (normalized )
curve the (special) unitary basis $E_{1},E_{2},\ldots ,E_{n+1}$ along the
curve ($\langle E_{i}|E_{j} \rangle = \delta_{ij}$). This new basis
(called the Frenet basis) satisfies the set of
linear equations
\beq \label{eq:Frenet}
\left( \ba{c} E_{1} \\ E_{2}  \\  \vdots \\  \vdots \\ E_{n+1} \ea \right)_{,z}
= \left(\ba{ccccc}
\theta^{1}_{,z} & \ee^{\theta^{2}-\theta^{1}} & 0 & \cdots & 0 \\
0 & \theta^{2}_{,z} & \ee^{\theta^{3}-\theta^{2}}&  & \vdots \\
  &  0 & \ddots & \ddots & 0 \\
\vdots &  & \ddots & \theta^{n}_{,z} & \ee^{\theta^{n+1}-\theta^{n}} \\
0 & \cdots &  & 0 &\theta^{n+1}_{,z} \ea \right)
\left( \ba{c} E_{1} \\ E_{2}  \\  \vdots \\  \vdots \\ E_{n+1} \ea \right)
\eeq
\[
\left( \ba{c} E_{1} \\ E_{2}  \\ \vdots  \\  \vdots \\ E_{n+1} \ea
\right)_{,\bar{z}} = - \left(\ba{ccccc}
\theta^{1}_{,\bar{z}}& 0 &  & \cdots & 0 \\
\ee^{\theta^{2}-\theta^{1}} & \theta^{2}_{,\bar{z}} & 0 &  & \vdots \\
 0   & \ee^{\theta^{3}-\theta^{2}}& \ddots & \ddots &  \\
\vdots  &  & \ddots & \theta^{n}_{,\bar{z}} & 0 \\
0   & \cdots & 0 & \ee^{\theta^{n+1}-\theta^{n}} &\theta^{n+1}_{,\bar{z}}
\ea \right)
\left( \ba{c} E_{1} \\ E_{2}  \\ \vdots   \\  \vdots \\ E_{n+1} \ea \right)
\; \; ,
\]
and the equations (\ref{eq:TodaA}) are compatibility conditions of the above
linear system.

\bigskip

\nin One can show \cite{DZ,EeWo} that all the vectors
$E_{k}\in \CC^{n+1}$ represent harmonic maps ($\sigma$-models in
physical terminology) from the Riemann surface ${\cal R}$
to the complex projective space $\CC\PP^{n}$ equipped with the standard
(Fubini -- Study) metric.

\bigskip

\nin Open Toda systems related to other simple non-exceptional Lie algebras are
various reductions of this general case. It turns out
that all the reduced holomorphic curves were studied using the twistor
techniques \cite{Bryant3}.

\section{Isotropic curves and $B_{m}$-Toda systems}

\label{sec:B}
Geometry of the $B_{m}$-Toda systems can be explained with the help of
the complex quadric
\beq
{\cal Q}_{2m-1} = \{ [z]\in\CC\PP^{2m} : \; (z|z)=\sum_{k=1}^{2m+1}z_{k}^{2} =
0 \} \; \; . \eeq
Holomorphic curves $\phi:{\cal R} \ra{\cal Q}_{2m-1}\subset
\CC\PP^{2m}$ are usually called isotropic curves.

\bigskip

\nin {\bf Definition 1} A nondegenerate holomorphic curve $\phi:{\cal R} \ra
\CC\PP^{2m}$ is called maximally isotropic if its $m$th associated curve
$\phi_{m}:{\cal R} \ra\GG(m,2m+1)$
is made out of isotropic $m$-planes (we write $\phi_{m}\subset{\cal
Q}_{2m-1}$).

\bigskip

\nin One can observe that $\phi_{m+1}\subset {\cal Q}_{2m-1}$ cannot hold for
nondegenerate curves. When $\phi$ is a maximally isotropic curve then for
$k=1,\ldots ,m$ all the associated curves $\phi_{k}$
are contained in the quadric ${\cal Q}_{2m-1}$. In terms of local holomorphic
lift $f:{\cal O}\ra\CC^{2m+1}$ this property is equivalent to
\beq \label{eq:izot}
(f^{(k)}|f^{(k)}) = 0 \; \; \; , \; \; k=0,\ldots ,m-1 \; \; \; . \eeq
The first $m$ vectors $E_{1}, \ldots ,E_{m}$ of the Frenet frame are
isotropic vectors what implies that together with $E_{m+1}$ and
$\bar{E}_{1},\ldots ,\bar{E}_{m}$ they form unitary frame along the curve.
Moreover, the equations satisfied by $\bar{E}_{k}$ can be found from those of
$E_{k}$, e.g.
\beq
\bar{E}_{k,z} = -\theta^{k}_{,z}\bar{E}_{k} -
\ee^{\theta^{k}-\theta^{k-1}}\bar{E}_{k-1}
\; \; .\eeq
If we impose the normalization
condition
\beq \label{eq:normB} (f^{(m)}|f^{(m)})=1 \eeq
then $E_{m+1}\in\RR\PP^{2m}$ (embedded totally geodesic in $\CC\PP^{2m}$)
\cite{EeWo,Cal} and the vectors
\beq E_{1},\ldots ,E_{m},E_{m+1},-\bar{E}_{m}, \ldots, (-1)^{k}\bar{E}_{m+1-k}
,\ldots ,(-1)^{m}\bar{E}_{1} \eeq
form the Frenet frame along the maximally isotropic curve.

\bigskip

\nin From the linear system (\ref{eq:Frenet}) we obtain that the functions
$\theta^{k}$ defined in (\ref{def:thetaA}) are now subjected to the condition
\beq \theta^{2m+2-k} = -\theta^{k} \eeq
which implies $\theta^{m+1}\equiv 0$ and equations (\ref{eq:TodaA}) reduce to
\bea \label{eq:TodaB}
2\theta^{1}_{,z\bar{z}} & = & \ee^{2(\theta^{2} - \theta^{1})} \nonumber \\
2\theta^{l}_{,z\bar{z}} & = & \ee^{2(\theta^{l+1} - \theta^{l})} -
 \ee^{2(\theta^{l} - \theta^{l-1})} \; \; , \; \; l=2, \ldots , m-1\nonumber \\
2\theta^{m}_{,z\bar{z}} & = & \ee^{ -2\theta^{m}}-
\ee^{2(\theta^{m} - \theta^{m-1})} \; \; . \eea
This matches with the standard representation of the simple roots
and Cartan subalgebra of $B_{m}\cong{\frak s}{\frak o}(2m+1)$
in terms of the orthonormal basis of $\RR^{m}$
\beq
\pi = \{ \eee_{1}-\eee_{2},\ldots , \eee_{m-1}-\eee_{m}, \eee_{m} \}
\cong \{ h_{1}, h_{2}, \ldots ,h_{m} \} \; \; .
\eeq

\medskip

\nin {\bf Theorem 2 \cite{DolSy2}} The Pl\"{u}cker formulas for maximally
isotropic holomorphic curves in $\CC\PP^{2m}$ form the open Toda system
related to the compact real form of ${\frak s}{\frak o}(2m+1)$ Lie algebra.

\bigskip

\nin One can show \cite{Yang} that for $k=1,\ldots ,m$ the vectors $E_{k}$
represent harmonic immersions of the Riemann surface ${\cal R}$
into the quadric ${\cal Q}_{2m-1}\subset\CC\PP^{2m}$ (with the metric structure
inherited from
the ambient complex projective space) and $E_{m+1}$ represents conformal
minimal
immersion of ${\cal R}$ in $\RR\PP^{2m}$, which lifted to $\SS^{2m}$ is called
pseudoholomorphic immersion \cite{Cal}.

\bigskip

\nin One should say also that all the immersions of Riemann
surfaces considered in this paper can have isolated branching points. Such
immersions are called branched immersions
\cite{GOR}. The induced metrics on the surface are in fact pseudometrics
\cite{Yang,GrHa}.

\bigskip

\nin The rest of this Section is devoted to parametrization of maximally
isotropic curves in terms of meromorphic functions. This provides us with the
geometric meaning of the arbitrary functions entering into general solutions of
the open Toda systems \cite{LeSav} and gives the geometric interpretation of
Liouville-B\"{a}cklund transformations for such systems \cite{Andr,Popowicz1}.

\bigskip

\nin It is convenient to use instead of the standard basis
$\{ \eee_{k}\}_{k=1}^{2m+1}$ of $\CC^{2m+1}$ the "maximally isotropic" unitary
basis
\beq \label{izotbasis}
\fff_{\pm k}=\frac{\eee_{2k-1} \pm i \eee_{2k}}{\sqrt{2}} \; \; \; \;
k=1,\ldots m
\; \; \; , \; \; \fff_{0}=\eee_{2m+1} \; \; .
\eeq
The new coordinates $\{ w_{k}\}_{k=-m}^{m}$ can be expressed in terms of the
old
ones $\{ z_{l}\}_{l=1}^{2m+1}$ as
\beq w_{\pm k} =\frac{z_{2k-1} \mp i z_{2k}}{\sqrt{2}} \;
\; \; \;  k=1,\ldots m  \; \; \; , \; \;  w_{0}=z_{2m+1} \; \;
\eeq
and the form $(\; |\; )$ which defines the quadric ${\cal Q}_{2m-1}$ reads
\beq
(z|z) = w_{0}^{2} + 2\sum_{k=1}^{m}{w_{-k}w_{k}} = {\cal F}(w) \; \; .
\eeq
When $w_{[1]}:{\cal O}\ra\CC^{2m+1}$ is a local holomorphic lift of the
maximally isotropic curve written in
terms of the new coordinates then the isotropy condition
\beq \label{eq:izot1} {\cal F}(w_{[1]}) = 0 \eeq
allows to reduce the number of functions necessary to describe the curve. We
can represent any isotropic curve in the form
\beq \label{eq:w1B}
w_{[1]} = u_{1}\left(1,w_{[2]},-\frac{1}{2}{\cal F}(w_{[2]})\: \right)
\; \; , \eeq
where (we denote by the same letter ${\cal F}$ the quadratic form in
$\CC^{2m-1}$)
\beq u_{1} = w_{[1]-m} \; \;  , \; \; w_{[2]k} =\frac{w_{[1]k}}{u_{1}}  \; \; ,
\; \; k = -m+1,\ldots , m-1 \; \; .
\eeq
Moreover one can show that
\beq {\cal F}(w'_{[1]})= u_{1}^{2}{\cal F}(w_{[2]}') \; \; ,
\eeq
and the next isotropy condition (see equation (\ref{eq:izot})) allows
to parametrize $w'_{[2]}$ in the form as above
\beq
w_{[2]}' = u_{2}\left(1, w_{[3]},-\frac{1}{2}{\cal F}(w_{[3]})\,\right)
\; \;  , \eeq
and
\beq {\cal F}(w''_{[1]})= u_{1}^{2}u_{2}^{2}{\cal F}(w_{[3]}') \; \; .
\eeq
This way for maximally isotropic curve one can define $m$ meromorphic functions
$u_{k}$ such that
\beq \label{eq:wkB}
w_{[k]}' = u_{k}\left(1, w_{[k+1]},-\frac{1}{2}{\cal F}(w_{[k+1]})\,\right) \in
\CC^{2m+3-2k}  \; \; \; \;  k=2,\ldots ,m \; . \eeq
Moreover the normalization condition (\ref{eq:normB}) implies
\beq \label{eq:normBw}
u_{1}^{2}u_{2}^{2}\cdot \ldots \cdot u_{m}^{2}v^{2} = 1 \; \; \;
, \; \; (v=w'_{[m+1]}\in\CC) \; \; . \eeq
One can reverse this process, and starting from $m$ meromorphic local functions
$u_{k}$ (and some constants of integration, which should play more important
role in the general case of complex Lie algebras) and using equations
(\ref{eq:w1B})(\ref{eq:wkB})(\ref{eq:normBw}) reconstruct locally the
normalized maximally isotropic curve, and then the corresponding solution to
the Toda system (\ref{eq:TodaB}). Details and examples can be found in
\cite{Doliwa}.

\bigskip

\nin The parametrization presented above can be considered also as a method of
construction of maximally isotropic curves in $\CC\PP^{2m}$ from such curves
in $\CC\PP^{2m-2}$ (see also \cite{BoGa} where $\sigma$-models with finite
action were constructed in a similar way). On the level of Toda systems this
process
corresponds to the Liouville--B\"{a}cklund transformation \cite{Andr} between
open $B_{m}$-system and $B_{m-1}$ system plus Laplace equation.

\section{Horizontal curves and $C_{m}$-Toda systems}

\label{sec:C}
Geometry of $C_{m}$-Toda systems can be described in terms of the celebrated
Penrose fibration
$ \CC\PP^{2m-1} \stackrel{\CC\PP^{1}}{\longrightarrow} \HH\PP^{m-1} $
(here we consider the right quaternionic projective space) obtained from the
standard identification of the quaternionic space $\HH$ with $\CC^{2}$
\beq \HH\ni q=a+jb \mapsto (a,b)\in \CC^{2} \; \; . \eeq
In terms of homogeneous coordinates
\beq \pi[(z_{1},z_{2})]_{\CC} = [z_{1} + jz_{2}]_{\HH} \; \; \; \; \; , \; \;
\; z_{1},z_{2}\in\CC^{m} \; \; . \eeq
The fiber over $\pi[(z_{1},z_{2})]_{\CC}$ is the projective line in
$\CC\PP^{2m-1}$ through $[(z_{1},z_{2})]_{\CC}$ and
$[(-\bar{z}_{2},\bar{z}_{1})]_{\CC} $
(note that
$ (z_{1}+jz_{2})j = -\bar{z}_{2}+j\bar{z}_{1} \; \; $).

\bigskip

\nin The complement to the fibre orthogonal with respect to the standard
Fubini-Study metric in $\CC\PP^{2m-1}$ defines the horizontal distribution
${\cal H}$.
The vector $v\in T_{[z]}\CC\PP^{2m-1}$ is horizontal if
\beq \langle (z_{1},z_{2})j|(v_{1},v_{2})\rangle =
\langle (-\bar{z}_{2},\bar{z}_{1})|(v_{1},v_{2})\rangle = (z_{1}|v_{2}) -
(z_{2}|v_{1}) = 0 \; \; . \eeq

\bigskip

\nin {\bf Definition 2} A non-degenerate holomorphic curve $\phi:{\cal R}\ra
\CC\PP^{2m-1}$ is called horizontal if it is tangent to the distribution ${\cal
H}$.

\bigskip

\nin In terms of a local holomorphic lift $f=(f_{1},f_{2}):
{\cal O}\ra \CC^{2m}$
this property is equivalent to
\beq \langle fj | f' \rangle = (f_{1}|f_{2}') - ( f_{2}|f_{1}') = 0 \; \; .
\eeq
One can generalize the notion of horizontality to the associated curves.

\bigskip

\nin {\bf Definition 3} A nondegenerate holomorphic curve  $\phi:{\cal R} \ra
\CC\PP^{2m-1}$ is called superhorizontal if its $(m-1)$th associated curve
$\phi_{m-1}:{\cal R} \ra\GG(m-1,2m)$ is horizontal.

\bigskip

\nin This implies that for $k=1,\ldots ,m-1$ all the curves $\phi_{k}$ are
also horizontal. In terms of a local holomorphic lift
it is equivalent to (compare with (\ref{eq:izot}))
\beq \label{eq:horiz}
\langle f^{(k)}j|f^{(k+1)}\rangle =
(f_{1}^{(k)}|f_{2}^{(k+1)}) - ( f_{2}^{(k)}|f_{1}^{(k+1)})= 0 \; \; \; ,
\; \; k=0,\ldots ,m-2 \; \; \; .  \eeq
The superhorizontal curve defines a special type of complex frame (called the
symplectic frame). Let us take the
first $m$ vectors $E_{1},\ldots ,E_{m}$ of the Frenet frame. The
superhorizontality of the curve implies that together with the vectors
$E_{1}j,\ldots ,E_{m}j$ they form unitary frame along the curve . Moreover, the
equations satisfied by $E_{k}j$ can be found from those of
$E_{l}$, e.g.
\beq
(E_{k}j)_{,z} = -\theta^{k}_{,z}E_{k}j - \ee^{\theta^{k}-\theta^{k-1}}E_{k-1}j
\; \; .\eeq
One can show that the vector $E_{m+1}$ of the natural unitary Frenet frame is
proportional to $E_{m}j$. The phase factor can be fixed by the following
normalization of the local holomorphic lift
\beq \label{eq:normC}
\langle f^{(m-1)}j|f^{(m)}\rangle = 1
\eeq
which implies $E_{m+1} = E_{m}j$.

\bigskip

\nin Putting all the facts together we obtain that the vectors
\beq
E_{1}, \ldots ,E_{m},E_{m}j, \ldots ,(-1)^{k}E_{m-k}j , \ldots
,(-1)^{m-1}E_{1}j \; \; \eeq
form the Frenet frame along the superhorizontal curve with the corresponding
functions
\beq \theta^{1}, \ldots ,\theta^{m},- \theta^{m},\ldots ,-\theta^{1} \; \;
\eeq
and equations (\ref{eq:TodaA}) reduce to
\bea \label{eq:TodaC}
2\theta^{1}_{,z\bar{z}} & = & \ee^{2(\theta^{2} - \theta^{1})} \nonumber \\
2\theta^{l}_{,z\bar{z}} & = & \ee^{2(\theta^{l+1} - \theta^{l})} -
 \ee^{2(\theta^{l} - \theta^{l-1})} \; \; , \; \; l=2, \ldots , m-1\nonumber \\
2\theta^{m}_{,z\bar{z}} & = & \ee^{ -4\theta^{m}}-
\ee^{2(\theta^{m} - \theta^{m-1})} \; \; . \eea
This matches with the standard representation of the simple roots and
Cartan subalgebra of $C_{m}\cong{\frak s}{\frak p}(m)$
in terms of the orthonormal basis of $\RR^{m}$
\beq
\pi = \{ \eee_{1}-\eee_{2},\ldots , \eee_{m-1}-\eee_{m}, 2\eee_{m} \}
\cong \{ h_{1}, h_{2}, \ldots ,h_{m} \} \; \; .
\eeq

\medskip

\nin {\bf Theorem 3} The Pl\"{u}cker formulas for superhorizontal
holomorphic curves in $\CC\PP^{2m-1}$ form the open Toda system
related to the compact real form of ${\frak s}{\frak p}(m)$ Lie algebra.

\bigskip

\nin One can show \cite{Yang} that the vectors of the symplectic frame
represent harmonic branched immersions of ${\cal R}$ into $\HH\PP^{m-1}$.

\bigskip

\nin The general solution to the system (\ref{eq:TodaC}) can be found from
parametrization of
the superhorizontal curves subjected to the normalization condition
(\ref{eq:normC}). When $f_{[1]}:{\cal O}\ra \CC^{2m}$ is a local holomorphic
lift of such a curve then the horizontality condition
\beq \langle f_{[1]}j|f_{[1]}'\rangle = 0 \eeq
allows to represent it (we arrange coordinates of a point $z\in\CC^{2m}$
in pairs) as
\beq f_{[1]}=u_{1}\left( 1 ,-\int \langle f_{[2]}j|f_{[2]}'\rangle dz ,f_{[2]}
\right) \; \; , \eeq
where
\beq u_{1} = f_{[1]1} \; \; , \; \; \; \; f_{[2]k} = \frac{f_{[1]k+2}}{u_{1}}
\;
\; \; \; \; k=1, \ldots ,2m-2 \; \; . \eeq
The next condition
\beq \langle f_{[1]}'j|f''_{[1]}\rangle = u_{1}^{2}\langle f'_{[2]}j|f_{[2]}''
\rangle = 0 \eeq
allows to represent $f_{[2]}'$ similarly.

\bigskip

\nin This way any superhorizontal curve defines $m$ meromorphic functions
$u_{k}$ such that for $k= 2 , \ldots m-1$
\beq f_{[k]}'=u_{k}\left( 1 ,-\int \langle f_{[k+1]}j|f_{[k+1]}'\rangle dz ,
f_{[k+1]} \right) \in \CC^{2m+2-2k}\; \; ,  \eeq
and
\beq f_{[m]}' = u_{m}(1,v)\in \CC^{2} \; \; . \eeq
The normalization condition (\ref{eq:normC}) implies
\beq u_{1}^{2} u_{2}^{2}\cdot \ldots \cdot u_{m}^{2}v' = 1 \; \; . \eeq
One can reverse this process and from $m$ meromorphic local functions
$u_{k}$ (and some constants of integration) reconstruct the
normalized local holomorphic lift of the superhorizontal curve, and then
the corresponding solution to the $C_{m}$-Toda system (\ref{eq:TodaC}).

\section{Isotropic curves and $D_{m}$-Toda systems}

\label{sec:D}
Geometric interpretation of the open $D_{m}\cong{\frak s}{\frak o}(2m)$ Toda
systems is similar to that of $B_{m}\cong{\frak s}{\frak o}(2m+1)$.
We start from nondegenerate holomorphic curve $\phi:{\cal R}\ra
\CC\PP^{2m-1}$ subjected to the maximal isotropy condition
\beq \phi_{m-1}\subset {\cal Q}_{2m-2} \; \; . \eeq
In terms of local holomorphic lift $f:{\cal O}\ra \CC^{2m}$ this condition is
equivalent to
\beq \label{eq:izotD} (f^{(k)}|f^{(k)}) = 0 \; \; \; , \; \;
k=0,\ldots ,m-2 \; \; \; . \eeq
Let us take the first $m-1$ vectors $E_{1}, \ldots ,E_{m-1}$ of the Frenet
frame
along the curve. These vectors  are isotropic, what implies that they are
orthogonal to their conjugate $\bar{E}_{1}, \ldots ,\bar{E}_{m-1}$. The next
vector $\tilde{E}_{m}$ of the Frenet frame
(we add the tilde for convenience) is orthogonal to the space $V$
spanned
by $E_{1}, \ldots ,E_{m-1},\bar{E}_{1}, \ldots ,\bar{E}_{m-1}$ and can be
decomposed into sum of a pair of complementary isotropic vectors
$E_{m},\bar{E}_{m}\in V^{\bot}$ (such a pair is given up to a phase factor).
When $\tilde{\theta}^{m}$ is the "old" function corresponding to
$\tilde{E}_{m}$ then we can write
\beq \ee^{\tilde{\theta}^{m}}\tilde{E}_{m}=a E_{m} + b \bar{E}_{m} \; \; . \eeq
We fix the phase factor by the condition
\beq a=\ee^{\theta^{m}}\in\RR_{+} \eeq
defining this way new local function $\theta^{m}$. The normalization condition
\beq \label{eq:normD} (f^{(m-1)}|f^{(m-1)}) = 2 \eeq
implies $b=a^{-1}$, and this way
\beq \label{cond:D} \ee^{\tilde{\theta}^{m}}\tilde{E}_{m}=
\ee^{\theta^{m}} E_{m} +
\ee^{-\theta^{m}} \bar{E}_{m} \; \; , \; \; \; \;
\ee^{2\tilde{\theta}^{m}}=2\cosh 2\theta^{m} \; \; .\eeq
The new vectors are subjected to the equations
\bea E_{m,z} & = & \theta^{m}_{,z}E_{m} -
\ee^{-\theta^{m}-\theta^{m-1}}\bar{E}_{m-1} \; \; , \nonumber \\
E_{m,\bar{z}} & = & -\theta^{m}_{,\bar{z}}E_{m} -
\ee^{\theta^{m}-\theta^{m-1}}{E}_{m-1} \; \; , \nonumber \\
\bar{E}_{m,z} & = & -\theta^{m}_{,z}\bar{E}_{m} -
\ee^{\theta^{m}-\theta^{m-1}}\bar{E}_{m-1} \; \; , \nonumber \\
\bar{E}_{m,\bar{z}} & = & \theta^{m}_{,\bar{z}}\bar{E}_{m} -
\ee^{-\theta^{m}-\theta^{m-1}}{E}_{m-1} \; \; . \eea
The following diagram describes the new unitary and isotropic frame (the arrow
denotes differentiation with respect to $z$)

\bigskip

{\scriptsize
\[
\left( \ba{c} \theta^{1}  \\ E_{1} \ea \right) \ra
 \ldots \ra \left( \ba{c}  \theta^{m-1} \\ E_{m-1}  \ea \right)
\ba{ccc}
 \nearrow & \left( \ba{c}\theta^{m} \\ E_{m} \ea \right)  & \searrow \\
 \searrow & \left( \ba{c} -\theta^{m} \\ \bar{E}_{m} \ea\right)  &\nearrow
\ea
\left( \ba{c}-\theta^{m-1} \\ -\bar{E}_{m-1} \ea \right) \ra \ldots
\left( \ba{c} -\theta^{m-k} \\ (-1)^{k}\bar{E}_{m-k} \ea \right) \ldots \ra
\left( \ba{c} -\theta^{1} \\ (-1)^{m-1}\bar{E}_{1} \ea \right)
\] }
and the new equations read
\bea \label{eq:TodaD}
2\theta^{1}_{,z\bar{z}} & = & \ee^{2(\theta^{2} - \theta^{1})} \nonumber \\
2\theta^{l}_{,z\bar{z}} & = & \ee^{2(\theta^{l+1} - \theta^{l})} -
 \ee^{2(\theta^{l} - \theta^{l-1})} \; \; , \; \; l=2, \ldots , m-2\nonumber \\
2\theta^{m-1}_{,z\bar{z}} & = & \ee^{2(\theta^{m} - \theta^{m-1})}
+ \ee^{-2(\theta^{m} + \theta^{m-1})}-
\ee^{2(\theta^{m-1} - \theta^{m-2})} \; \; , \nonumber \\
2\theta^{m}_{,z\bar{z}} & = & - \ee^{2(\theta^{m} - \theta^{m-1})}
+\ee^{-2(\theta^{m} + \theta^{m-1})} \; \; . \eea
This matches with the standard representation of the simple roots and Cartan
subalgebra of $D_{m}\cong{\frak s}{\frak o}(2m)$
in terms of the orthonormal
basis of $\RR^{m}$
\beq
\pi = \{ \eee_{1}-\eee_{2},\ldots , \eee_{m-1}-\eee_{m}, \eee_{m-1}+\eee_{m} \}
\cong \{ h_{1}, h_{2}, \ldots ,h_{m} \} \; \; .
\eeq

\medskip

\nin {\bf Theorem 4} The Pl\"{u}cker formulas (modified by conditions
(\ref{cond:D})) for maximally isotropic
holomorphic curves in $\CC\PP^{2m-1}$ form the open Toda system
related to the compact real form of ${\frak s}{\frak o}(2m)$ Lie algebra.

\medskip

\nin Parametrization of the maximally isotropic curves in $\CC\PP^{2m-1}$ is
similar to that in $\CC\PP^{2m}$. The only difference (except for the fact that
we use now the fully isotropic basis in $\CC^{2m}$ dropping the vector
$\fff_{0}$, see (\ref{izotbasis})) is at the last step of the construction.
To show this it is enough to consider a simple case, say $m=3$.

\bigskip

\nin The isotropic curve $\phi:{\cal R}\ra {\cal Q}_{4}\subset \CC\PP^{5}$ can
be represented locally in isotropic coordinates as
\beq\label{eq:w1D} w_{[1]}= u_{1}\left( 1 ,w_{[2]},
-\frac{1}{2}{\cal F}(w_{[2]})\: \right) \in \CC^{6}
\; \; , \eeq
where $w_{[2]}:{\cal O}\ra \CC^{4}$. The next isotropy condition implies that
also
\beq
w_{[2]}' = u_{2}\left(1, w_{[3]},-\frac{1}{2}{\cal F}(w_{[3]})\,\right)
\; \; , \eeq
and the normalization condition (\ref{eq:normD}) implies
\beq \label{eq:normDw}
u_{1}^{2}u_{2}^{2}u_{3}^{2}v = 1 \; \; \; , \; \; ( w_{[3]}'=u_{3}(1,v) \in
\CC^{2}) \; \;  . \eeq
The details are left to the Reader.

\section{Conclusion}

Geometry of special holomorphic curves in relation
to the open Toda
systems for non-exceptional simple Lie algebras was described. The legs of the
Frenet frame along the curves give rise to (pseudoholomorphic) harmonic maps
into complex projective spaces, quadrics, Euclidean spheres and quaternionic
projective spaces.

\medskip

\nin Parametrization of the curves under consideration, which generalizes
the Weierstrass representation for minimal surfaces in $\EE^{3}$ in terms of
meromorphic functions was also found. This provides with the
geometric origin of the general solution to the open Toda systems and their
Liouville--B\"{a}cklund transformations.

\medskip

\nin Also the known reductions of the $A_{n}$ systems to $B_{m}$ and $C_{m}$
systems were explained on the geometric level. Moreover, embedding of
the $D_{m}$ Toda system into $A_{2m-1}$ system was given.

\bigskip

\nin In the forthcoming paper \cite{Doliwa2} it will be shown how periodic Toda
systems (which are genuine soliton systems) can be also described in relation
to harmonic maps of Riemann surfaces.

\bigskip

\bigskip

\nin {\Large {\bf Acknowledgments}}

\bigskip

\bigskip

\nin I would like to thank Prof. Sym for suggesting the problem
and Prof. Popowicz for pointing out reference \cite{GervMats}.

\end{document}